# Co-design of a novel CMOS highly parallel, low-power, multi-chip neural network accelerator

W. Hokenmaier, R. Jurasek
*Semiconductor Hardware Development*
Green Mountain Semiconductor Inc.
Burlington, VT USA
whokenmaier@greenmountainsemi.com
rjurasek@greenmountainsemi.com

E. Bowen, R. Granger, D. Odom
*Software, Algorithm, and Architecture Design*
Non-Von LLC
Lyme, NH USA
Eli@non-von.com
richard_granger@non-von.com
DerekOdom@non-von.com

*Abstract*—Why do security cameras, sensors, and siri use cloud servers instead of on-board computation? The lack of very-low-power, high-performance chips greatly limits the ability to field untethered edge devices. We present the NV-1, a new low-power ASIC AI processor that greatly accelerates parallel processing ($\sim$10X) with dramatic reduction in energy consumption ($>$ 100X), via many parallel combined processor-memory units, i.e., a drastically non-von-Neumann architecture, allowing very large numbers of independent processing streams without bottlenecks due to typical monolithic memory. The current initial prototype fab arises from a successful co-development effort between algorithm- and software-driven architectural design and VLSI design realities. An innovative communication protocol minimizes power usage, and data transport costs among nodes were vastly reduced by eliminating the address bus, through local target address matching. Throughout the development process, the software/architecture team was able to innovate alongside the circuit design team's implementation effort. A digital twin of the proposed hardware was developed early on to ensure that the technical implementation met the architectural specifications, and indeed the predicted performance metrics have now been thoroughly verified in real hardware test data. The resulting device is currently being used in a fielded edge sensor application; additional proofs of principle are in progress demonstrating the proof on the ground of this new real-world extremely low-power high-performance ASIC device.

## I. Introduction

The low-power, high-performance AI processor described here, the NV-1, is a joint development effort between Non-Von LLC and Green Mountain Semiconductor. Non-Von's architecture designs originated from a novel machine "instruction set" of fundamental parallel operations and software, that were initially derived (at Dartmouth College's Brain Engineering Laboratory) from the operation and arrangement of circuitry in the brain [1]. The collaboration between Non-Von and Green Mountain Semiconductor then arose to confront the challenge of translating this instruction set into a correspondingly efficient hardware implementation. Throughout the collaborative process, from the initial architecture designs all the way through tapeout, a digital-twin approach has been used to enable the closed-loop communication between hardware capabilities from the engineering team and algorithm developments from the software team. The process included sometimes modifying or even dropping certain instructions in order to keep the node size minimal. This effective simulation and communication approach facilitated the two teams' joint optimization of the final design for size, performance, and power, ensuring that proposed hardware implementations continued to meet the intended performance targets.

Notably, the approach encompasses the design of allowing AI network sizes far beyond a single die; the communication protocol expands seamlessly beyond individual die boundaries, allowing a multitude of identical chiplet processors to be connected to achieve a targeted network node count. Thus a given configuration may be as small as a single chip or chiplet for applications domains such as internet-of-things (IoT) low-power devices, and also can directly scale up to huge arrays for some uses such as server farms, while still operating at comparatively very low power budgets. (Although the approach is fully compatible with very small fab technology, the initial low-cost prototype presented here used 28nm TSMC manufacturing.) The interface integrates with FPGAs and SoCs for overall communication, so that one or multiple chiplets can act either alone or as massively parallel AI coprocessors.

The first prototype (NV-1) includes 3200 cores per chip with seamless I/O compatibility to increase array size via chaining chips. During testing, this chip achieved 447 GB/s per 0.25 W, thus demonstrating both high performance and a radical power-use improvement over other comparable hardware devices (see further discussion in Results). The chip also has been fielded in real-world settings, performing real-time processing of a chemical sensor, with a power budget of $<$ 10 mW, providing a direct initial demonstration that the chip is operational and applications-ready.

## II. Background

Software developers have forever been at the mercy of the hardware that is available to them [2]. The limitations of given hardware designs superimpose substantial constraints on algorithm and software design. In particular, algorithms that are intrinsically parallel will be enormously slowed down by typical hardware. The typical approach has been to use GPUs and related hardware, but GPUs were of course designed for specialized image-processing operations, rather than broader parallel algorithms, and most systems typically must be written (or re-written) for GPU compatibility.

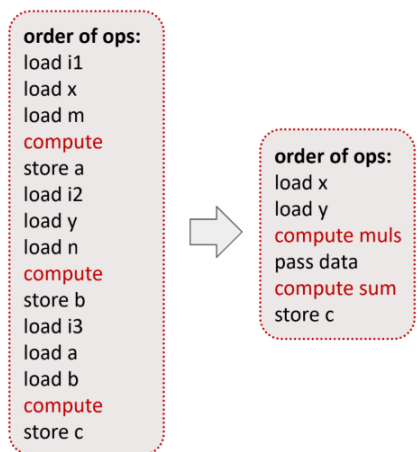

Fig. 1. Traditional instruction and reduced instructions for same task

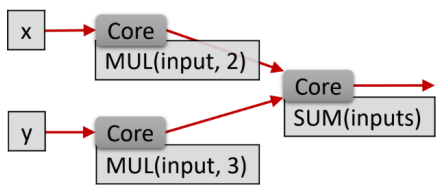

Fig. 2. Multiple cores executing instructions

The systems designed at Dartmouth and Non-Von [1] derive from operations of extremely large numbers of simple parallel elements in complex arrangements (neurons in brain circuitry); these are intrinsically massively parallel (rather than parallelized versions of inherently serial methods). Such intrinsically parallel algorithms are greatly sped up by appropriate parallel hardware, but it is highly rare and unusual for such hardware to be constructed for these parallel algorithms. Instead, the software must typically be adapted and compromised to available hardware, rather than new hardware architectures being developed to accommodate the parallel designs. The repurposing of GPUs to run accelerated neural networks provides this necessity for compromised software [3]; this approach is now so widespread that it is almost forgotten that GPUs are indeed far from an ideal hardware environment for parallel systems in general.

Again, GPUs were developed for particular image-processing tasks rather than for parallel algorithms in general. GPUs simply have been used in this adapted form solely because they existed, and they were far closer to parallel software needs than standard CPU designs. But to take seriously the needs of massively parallel software, and to design hardware specifically for these needs, has been almost entirely absent from the field. Moreover, the need for very low power use — such as required for fielded "internet of things" (IOT), sensors, medical devices, and much more, in environments where large batteries or power sources are extremely limited — has been a longstanding unmet need. Rather than the continued repurposing of hardware developed for other tasks, such as GPUs, the hardware presented here was specifically developed for low-power, high-performance massively parallel systems.

Current hardware for neural network solutions still use GPUs for training [4]. Hardware engineers have opened up low level access for software engineers to explore more efficient algorithms, optimizing data movement and improving efficiency. Because of the success of GPUs in neural network accelerations, engineers have developed many different hardware solutions from training chips, inference chips, low power edge devices and high performance cloud architectures. For example, Google has released the TPU (Tensor Processing Unit) for its data servers.

However, the current solutions for generative AI are not scalable, with current models taking racks containing hundreds of chips to run [4]. Moreover, the power needs (and cooling needs) for current hardware typically entails very specific siting for server farms, often specifically at sites of hydroelectric dams and other resources [5]. Convolutional neural networks, other deep neural networks, and transformers, all can be made somewhat more efficient, but for many hardware solutions it requires a large amount of batching to achieve efficiency.

From a software engineers perspective, this is clearly a limitation and drives solutions to an outcome that may not be needed for the original problem. Current state of the art instruction sets also impose limitations. Creating hardware that focuses specifically on the instructions necessary, instead allows for vastly more efficient designs to be realized. Most current cores have the ability to run in a flexible manner supporting more than a program may need [17], and although this gives a flexible processing architecture, it trades that flexibility for impaired performance. In NV-1 we instead focus on a specific instruction set that is hugely accelerated, while other portions of software can be picked up by a coprocessor.

### III. Design

Co-design of the software and hardware systems was crucial for rendering Non-Von's initial pioneering instruction set architecture for neural network acceleration into a complete working solution. To facilitate this parallel development of software and hardware, a digital twin was created in the form of a C++ software executable hardware model. This was done from the beginning of the project based initially on behavioral Verilog models, and then maintained throughout the project, as high level models were subsequently replaced with synthesized RTL code. The model allowed for the abstraction of hardware details and provided an equivalent behavioral representation of the hardware to be developed. By both parties agreeing on the functionality of the model, a clear goal for the programming and for the hardware design was defined. This methodology was the groundwork for later verification on the resulting hardware. Post tapeout, the results wanted from the silicon were to get physical power numbers along with showing functionality in hardware. The same waveforms used to simulate the chip were able to be used as vectors in the physical testing, further gaining confidence in the methodology.

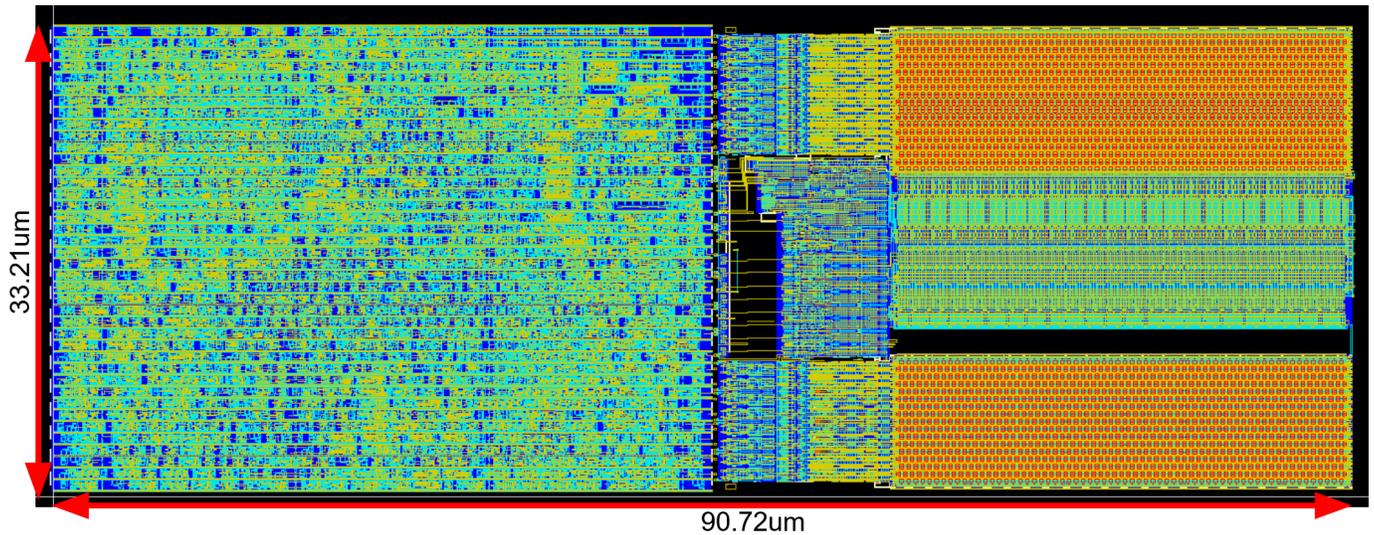

Fig. 3. Physical layout of a single core

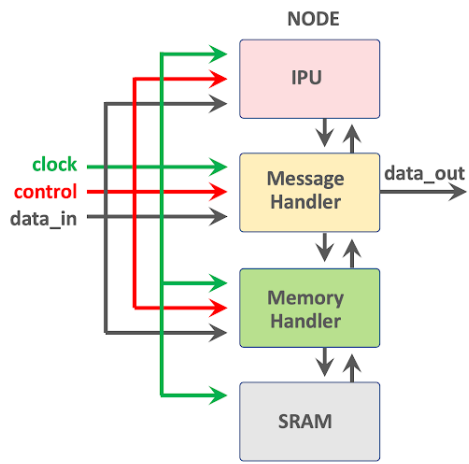

Fig. 4. Node sub-blocks

The architecture of a single NV-1 node is made up of four main sub-blocks. (Fig 4). The Message Handler is the interface of the node. The block handles all node-to-node communication on the bus, along with control of the system, decoding the nodes programming and initiating the system to start during a node activation. The Memory Handler and the SRAM work hand in hand, holding the nodes' communication information. The brains of the system is the IPU which handles the functionality of the nodes doing all of the calculations with data handed to it from the Message Handler. This structure is repeated throughout the whole chip in an array creating a distributed computational system that can process inferences with radically less power and fewer operations than in typical von Neumann implementations.

The initial minimal concept utilizes 64k cores. While any core can perform any of the defined instructions, in typical practice each core is initialized to perform just one task. By allowing only one task per core, the run time sending of instructions is not needed, and both the power and time for sending the instruction is removed. This is both different from a traditional CPU where instructions are sent for each command during execution, and from a GPU where a single instruction is sent to all cores and the same instruction is processed on every core with different data (SIMD). In the NV-1 chip presented here, data can be sent from each core to every other core. Each core maintains a boot-loaded address table defining its connections to other cores. This in particular was a concept easily realized in software, but not a straightforward task in hardware. Physical wiring limitations and timing considerations are problematic for bidirectional communication of 64k cores. Each core has a memory depth for core connections. 256 individual 16 bit numbers allow for the node to receive up to 256 other nodes output. An epoch is defined as the action of every core processing the messages from every other core in its received address memory and passing the results on for the next epoch. With intelligent programming of each core, repetitive tasks can be executed with very high efficiency.

For this prototype, a multi-project wafer tapeout, the maximum chip size was intentionally limited. The jointly developed reduced instruction set made it possible to optimize the core physical size to maximize the number of compute cores per die. Furthermore, innovation was needed to achieve a fully configurable bidirectional communication solution for up to 64k cores. The predefined address table removes the power and area intensive address bus, such that only data is being transmitted. This first prototype includes 3200 cores. It is notable that the communication protocol extends seamlessly beyond die boundaries, enabling the creation of arbitrary-

**Compute core utilization under memory bottleneck**

| | | |
|---|---|---|
| Non-Von NV1, single-chip configuration | 100% | [see derivations in this manuscript] |
| Embedded CPU, ARM Cortex-A8 | 50.8% | memory: DDR3 specs, TOPS: 2 DMIPS/MHz (x/1M fr MIPS to TIPS), x*1000 fr 1 MHz to 1 GHz, https://www.ti.com/lit/ds/symlink/am3358.pdf |
| NVIDIA Jetson TX2 | 0.73% | memory: https://developer.nvidia.com/embedded/jetson-tx2, TOPS from https://developer.nvidia.com/embedded/jetson-modules |
| NVIDIA Jetson Orin Nano 4GB | 0.06% | memory: https://developer.nvidia.com/embedded/jetson-modules, TOPS from https://tinyurl.com/NvidiaJetsonTops |
| Data ctr GPU, NVIDIA H100 SXM, tensor cores | 0.03% | memory: https://www.nvidia.com/en-us/data-center/h100/, TOPS from https://www.nvidia.com/en-us/data-center/h100/ |
| Google Coral Dev Board Micro | 0.03% | memory: LPDDR4 "(4-channel, 32-bit bus width)", https://tinyurl.com/CoralMem, int8 TOPS: https://coral.ai/products/accelerator-module/ |
| Google TPUv4 | 0.07% | https://cloud.google.com/tpu/docs/system-architecture-tpu-vm |
| Intel Habana Gaudi 2 | 0.63% | https://developer.habana.ai/resources/habana-models-performance/ |
| Tenstorrent Grayskull | 0.01% | https://tenstorrent.com/cards/ |
| Cerebras | 100% | https://f.hubspotusercontent30.net/hubfs/8968533/WSE-2%20Datasheet.pdf |
| Rebellions_Atom | 0.03% | https://rebellions.ai/wp-content/uploads/2023/09/Rebellions_ATOMProduct_Brief_v.3.2.pdf |
| Graphcore Colossus MK2 | 0.03% | memory from https://www.graphcore.ai/products/ipu, TOPS from https://www.graphcore.ai/products/ipu |

Fig. 5. Utilization percentages in the presence of memory bottlenecks

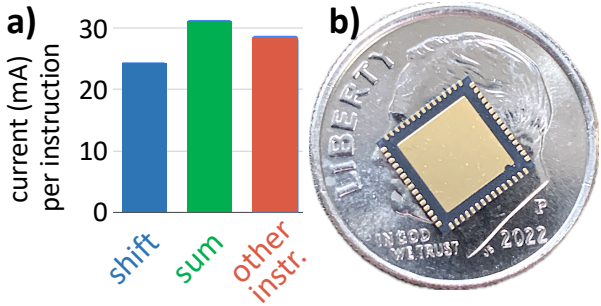

Fig. 6. (a) Relative current per instruction for NV-1 chip; (b) NV-1 (28nm TSMC fab)

TABLE I
CROSS-CHIP SLOPE INTERCEPT CURRENT AVERAGES(MA) ACCORDING TO FREQUENCY(MHZ)

| Condition | Slope |
|---|---|
| DIN at VSS: | Y = 3.25x + 6.3 |
| DIN at DVDD: | Y = 3.23x + 6.4 |
| DIN at ¼ Clk: | Y = 5.10x + 6.4 |
| DIN at ½ Clk: | Y = 6.95x + 6.4 |

sized arrays. Each die acts as a fully modular chiplet entity which can interact either with identical neighbors or can connect to a host computer or a hub which may in turn interface with other NV chiplet networks. Up to 21 chiplets can be combined to create an network of 64k cores. The first demonstration uses printed circuit board interconnects. Next generation designs target a significantly larger overall network size in the millions, and may leverage advanced high density 2.5D and 3D heterogeneous packaging methods for lowest power and further increased performance.

The prototype chip NV-1, a proof of concept array of 3200 nodes, was successfully completed via joint development efforts of GMS and Non-Von. The chip has functionality and showcases the architecture's very low power consumption. Designed in 28nm technology, the total array has dimensions of 3mm by 4mm. Further iterations of this device along with smaller technology nodes will continue to push a smaller footprint. Figure 3 shows the single node architecture, with its processing portion on the left and the SRAM block to save connectivity on the right. The digital twin throughout the design stages served as a blueprint for the design, ensuring that at each stage the hardware interpretation of what the network should be achieving lines up with the software concepts. This relationship continues on in the following section going past design and into verification, where the model is used to determine correct functionality in real time in silicon.

## IV. RESULTS

Throughout the design process the functionality of the chip was under the scrutiny of the Universal Verification Methodology (UVM). The expected data for this testbench was validated from both the GMS side and the Non-Von design teams. This proved to be a good vehicle of cross understanding from both hardware and software sides of the GMS and Non-Von design team. A shared C++ model was used to generate the expected data; this model was iteratively updated and checked by both teams to ensure that correct functionality was interpreted in the same way from the top level abstraction to the hardware. Once the correct functionality was agreed upon, the checker component of the UVM testbench could be utilized.

The testbench is able to run a full chip simulation in Verilog, with either random nodes or pre-programmed in order to test potential corner cases. The whole system is then run, both testing the proper setup procedure, and end-functionality correctness. The chip is viewed as a black box at the top to ensure proper data out and the nodes are also checked at the greybox level to ensure proper node-to-node communication. Within the testbench, nodes have been verified for correct message receiving and computation. This node message is then properly shifted through chip output and deemed correct at the black box level.

The verification effort found correct functionality for all of the instructions in the instruction set, along with correct communication between nodes, and proper operation at the chip level.

Figure 6a shows relative current per instruction for the NV-1 chip design, measured at 6.25 MHz, providing the root values for calculating speed and power tradeoffs, which are shown in Figures 5 and 7. It is worth noting that these figures amount to a max memory bandwidth of 447 GB/s per 0.25 W of power (number of nodes * single read per clock * clock speed, i.e., 447 GB/s = 3200 nodes * 50 MHz * (16 + 8 bits) / 8 / 1024 / 1024 / 1024) for a single NV-1 chip, and a corresponding 7.2 TB/s for an array of 16 chips. (Note that Fig 6 shows values

| | NV1 (1 chip) | NV1 (16 chips) | NV2 12nm (8x8mm chip) | NV2 7nm (6x6mm chip) | Embedded CPU ARM Cortex-A8 | NVIDIA Jetson TX2 | NVIDIA Jetson Orin Nano 4GB | NVIDIA H100 SXM (tensor cores) | Google Coral Dev Board Micro | Google TPUv4 |
|---|---|---|---|---|---|---|---|---|---|---|
| **::: Power (mW) :::** | | | | | | | | | | |
| Power, Idle | 6.2 | 99 | 18 | 10 | 17 | ~100 | ? | ? | 388 | 90,000 |
| Power, Nominal | 36 | 576 | 336 | 58 | ? | 7100 | ? | ? | 1050 | 170,000 |
| Power, Peak Workload | 243 | 3893 | 20,348 | 3091 | 1552 | 7500 | 10,000 | 700,000 | 3000 | 192,000 |
| **::: Adjusted Power* (mW @ 7 nm equivalent) :::** | | | | | | | | | | |
| Idle | 0.4 | 6.2 | 6 | 10 | 0.2 | ? | ? | ? | ? | ? |
| Nominal | 2.25 | 36 | 114 | 58 | ? | 1359 | ? | ? | ? | ? |
| Peak Workload | 15 | 243 | 6924 | 3091 | 18 | 1436 | 7656 | 2,143,750 | ? | ? |
| **::: Peak Compute Throughput (TOPS) :::** | | | | | | | | | | |
| Unstructured Sparse Data @ 50% | 0.2 | 2.6 | 41 | 67 | 0.002 | 1.3 | 10 | 1979 | 4 | 275 |
| Bool Arithmetic | 21 | 329 | 10,441 | 17,043 | 0.5 | ? | ? | ? | ? | ? |
| **::: Best-case Efficiency (TOPS/W) :::** | | | | | | | | | | |
| Unstructured Sparse @ 50% | 0.66 | 0.66 | 7 | 21 | 0.001 | 0.2 | 1 | 3 | 2 | 1.4 |
| Bool Arithmetic | 85 | 85 | 1908 | 5495 | 0.3 | ? | ? | ? | ? | ? |
| **::: Best-case Adjusted Efficiency** (TOPS/ adj W) :::** | | | | | | | | | | |
| Unstructured Sparse @ 50% | 11 | 11 | 6 | 22 | 0.1 | 1 | 1.3 | 1 | ? | ? |
| Bool Arithmetic | 1352 | 1352 | 1508 | 5513 | 28 | ? | ? | ? | ? | ? |

\* Power numbers, adjusting for differences in fab process; y = (nm^2) / (7^2)
\*\* TOPS per adjusted power

Fig. 7. Power, TOPS, and efficiency across multiple architectures

for 6.25 MHz whereas Fig 7 and memory bandwidth figures are values for 50 MHz).

The NV architecture was designed from first principles to eliminate almost all memory bandwidth bottlenecks, which is a considerable throughput and efficiency limitation in CPUs and GPUs. Because it is so typical for memory to be off-chip, the concept of memory bandwidth is thus often thought of in terms of I/O protocol (such as DDR3), rather than in terms of the effect that it has on the time and efficiency costs of real applied usage. Imagine beginning with a current GPU and inquiring how its performance would be affected by changes to its memory. First of all, if memory could be placed on-chip this would itself result in an enormous speedup in processing of the GPU in real applications. Even with on-chip memory, much of the von-Neumann bottleneck would still slow the system down if that memory still has to be treated as a monolithic entity that must be processed, so secondly, if the newly on-chip memory could then instead be distributed across processing units into memory blocks that were independent of each other, then further speedups could be achieved. These two steps (placement on chip, and independent distribution across processors) are at the heart of the new architecture, rendering it highly non-von-Neumann in design.

Note that these enormous speedups do not change the TOPS measures at all. TOPS measures are treated independently of any memory usage costs. That is, enormous speedups due to elimination of memory bottlenecks will not even show up as an improvement, if all one looks at are TOPS measures. Thus TOPS measures are highly misleading in such cases, since they cannot reflect speedups that arise due to re-architecting of memory.

We therefore provide a range of measures that are intended to enable approximate apples-to-apples comparisons, i.e., what theoretic and pragmatic gains would be achieved when switching from the characteristics of one type of chip to another type, such as CPUs to GPUs, CPUs or GPUs to non-von-Neumann architectures, etc.

A contemporary GPU has a reported peak memory bandwidth of 3.35 TB/s [4]. Calculating the peak memory bandwidth of NV-1 entails summing node-internal memory reads that can be performed during the course of computing a single operation: $f = (max\_num\_ops\_per\_sec * max\_bits\_per\_op) / 8 * 1024 * 1024 * 1024)$. Here we simply report the percent utilization that is possible given the nature of a memory bottleneck on particular hardware. Let $f = min(compute, bandwidth/n\_bytes\_per\_op) / compute$ where $n\_bytes\_per\_op = 3*16/8 = 6$ assumes that an operation uses two 16-bit inputs as operands and one 16-bit instruction. Then units(f) = ((GB/s / 1024) / bytes required per op) / TOPS. Figure 5 shows this as compute core utilization in the presence of the memory bottleneck.

We emphasize that these numbers are intended to illustrate the struggle that is presented by monolithic external memories. In practice, caches are used to avoid this, and those caches are not represented in these numbers in Fig 5. Standard approaches become very limited, as seen in the figure, because though they readily add more compute power (TOPS), they nonetheless cannot add memory bandwidth anywhere near as easily. (This is reflected in how the ARM Cortex does well in this figure: it is a single core, so not much compute to consume memory cycles.) In sum, this is not to say that memory bandwidth considerations are the sole factor in performance, but we wish to emphasize that it is in fact important and it is routinely overlooked in measures such as TOPS.

It should also be noted that the NV-1 is merely the first fabbed issuance of the Non-Von chip line; substantial further increases already are estimated in the upcoming NV-2 chip, using the same estimation methods that correctly led to previous very accurate predictions of NV-1 performance. It is highly notable that NV-1 does not use caches at all, nor a global memory space. Designers of GPUs extensively use caches to minimize the burden of their memory bottlenecks; these come at a cost of power, space (e.g., for cache coherence logic), and unpredictable timing. Figure 7 contains partial information, extracted from a range of sources, to roughly compare power, TOPS, and the resulting efficiency ratio, across a range of multiple different hardware architectures.

## V. Conclusions

The NV-1 test chip has been successfully manufactured (28nm TSMC technology), received in packaged dies, and functionally characterized and verified. System-level integration has been carried out to incorporate the chips in an existing sensor apparatus that has been tested in fielded conditions. The measured results from this new chip, shown in Figures 5, 6, and 7, demonstrate that it exhibits very high memory bandwidth performance at radically low power usage, outperforming standard competing chips by orders of magnitude.

The aim was to produce a new generation of AI hardware, rather than ongoing adaptation of systems such as GPUs, that were intrinsically designed for quite different purposes. The new NV platform is specifically designed to accelerate massively parallel software, thus providing a natural processor and coprocessor setting for innovative development of radically parallel systems. Moreover, these new platforms will execute at extremely low power — that is, at just tiny fractions of the power budgets of typical extant devices. Working demonstrations have been implemented to run the Whisper transformer-based real-time speech-to-text system with very low power, and to run a fielded real-time chemical sensor also with very low power ($< 10$ mW).

This project successfully demonstrates how software and hardware engineers can work together to co-design and optimize overall outcomes in terms of die size, performance, and power consumption. Rather than the necessity of compromising, via the use of hardware designs that happen to be there for other purposes, the possibility now arises to take innovative algorithms and software, and produce hardware ASIC designs that are well fitted to executing such software both with high performance and very low power.

With the ever increasing demands of AI hardware capabilities, especially in fielded low-power settings, this type of codevelopment effort, aided by a digital twin allowing for a continuous interdisciplinary verification and communication loop, may guide future projects to optimize TOPS/W not only as a pure hardware engineering task but as a joint endeavor. Design efforts are under way towards the next version, NV-2, which will further improve on power usage and minimize the physical size of each core through resource sharing.

Current edge-focused processors are highly challenged by restrictive low power budgets and high performance requirements at the edge in practice, and they still typically resort to using cloud computation that is costly (both in dollars and in power usage). We show here that even this initial prototype NV-1 device already drastically outperforms current technology in parallel computation tasks, both in performance and in power consumption. The ongoing approach addresses a very clear need that is seen across industries attempting to deploy AI and ML in real fielded applications.

## References

[1] Granger R. "Engines of the brain: The computational instruction set of human cognition." AI Magazine 27: 15-32 (2006).
  Moorkanikara J, Felch A, Chandrashekar A, Dutt N, Granger R, Nicolau A, Veidenbaum A. "Brain-derived vision algorithm on high-performance architectures." Int'l Journal of Parallel Prog. 27: 345-269 (2009).
  Chandrashekar A, Granger R. "Derivation of a novel efficient supervised learning algorithm from cortical-subcortical loops." Front. Comput. Neurosci. 5: 50. doi: 10.3389/fncom.2011.00050 (2012).
  Bowen E, Granger R, Rodriguez A. "A logical re-conception of neural networks: Hamiltonian bitwise part-whole architecture." Amer. Assoc. of Artif. Intell. (AAAI) https://openreview.net/forum?id=hP4dxXvvNc8 (2023).

[2] Barrett R, Borkarb S, Dosanjh S, Hammond S, Heroux M, Hu X, Luitjense J, Parker S, Shalf J, Tangd L. "On the role of co-design in high performance computing" Advances in Parallel Computing 24:141-155 (2013)

[3] Alcorn, P. "Intel 13th-Gen Raptor Lake specs, release date, benchmarks, and more" (20 Oct 2022) tinyurl.com/raptorlakespec

[4] https://www.nvidia.com/en-us/data-center/h100/
https://www.researchgate.net/publication/224262634_GPUs_and_the_Future_of_Parallel_Computing

[5] https://www.theatlantic.com/technology/archive/2024/03/ai-water-climate-microsoft/677602/
https://arxiv.org/abs/2312.12705

[6] memory: DDR3 specs, TOPS: 2 DMIPS/MHz (x/1M fr MIPS to TIPS), x*1000 fr 1 MHz to 1 GHz, https://www.ti.com/lit/ds/symlink/am3358.pdf

[7] memory: https://developer.nvidia.com/embedded/jetson-tx2, TOPS from https://developer.nvidia.com/embedded/jetson-modules

[8] memory: https://developer.nvidia.com/embedded/jetson-modules, TOPS from https://tinyurl.com/NvidiaJetsonTops

[9] memory: https://www.nvidia.com/en-us/data-center/h100/, TOPS from https://www.nvidia.com/en-us/data-center/h100/

[10] memory: LPDDR4 "(4-ch 32-bit bus width)", https://tinyurl.com/CoralMem,
int8 TOPS: https://coral.ai/products/accelerator-module/

[11] https://cloud.google.com/tpu/docs/system-architecture-tpu-vm

[12] https://developer.habana.ai/resources/habana-models-performance/

[13] https://tenstorrent.com/cards/

[14] https://tinyurl.com/cerebrasWSE-2

[15] https://tinyurl.com/rebellionsATOM

[16] memory from https://www.graphcore.ai/products/ipu, TOPS from https://www.graphcore.ai/products/ipu

[17] https://oscarlab.github.io/papers/instrpop-systor19.pdf

[*] Partial funding for the work reported herein was provided by the Office of Naval Research.